\documentstyle[12pt]{article}

\begin{document}
\thispagestyle{empty}
\rightline{gr-qc/9504008}
\rightline{MPI-PhT/95-28}
\rightline{April 1995}
 \begin{center}{
{\bf Non-Universality of Critical Behaviour\\
              \vglue 10pt
in Spherically Symmetric Gravitational Collapse
               }
\vglue 5pt
\vglue 1.0cm
{DIETER MAISON}\\
{\it Max-Planck-Institut f\"ur Physik, Werner-Heisenberg-Institut}\\
{\it F\"ohringer Ring 6, 80805 Munich, Germany}\\
\vglue 0.8cm
{ABSTRACT}
}\end{center}
\vglue 0.3cm
{\rightskip=3pc  \leftskip=3pc
\tenrm\baselineskip=12pt
\noindent
The aim of the present letter is to explain the `critical
behaviour' observed in numerical studies of spherically symmetric
gravitational collaps of a perfect fluid.
A simple expression results
for the critical index $\gamma$ of the black hole mass considered as an
order parameter.
$\gamma$ turns out to vary strongly with the parameter $k$ of the
assumed equation of state $p=k\rho$.
\vglue 0.8cm}
\noindent
Numerical simulations of the spherically symmetric gravitational
collapse of a massless scalar field
\cite{Chop}
have revealed a kind of critical behaviour with the mass of an
eventually formed black hole playing the role of an order parameter.
More precisely, Choptuik found a relation of the form
\begin{equation}
M_{\rm BH}\sim |p-p^*|^\gamma\;.
\label{crit}
\end{equation}
where $M_{\rm BH}$ is the mass of the formed black hole, $p^*$ is the
critical value of a parameter $p$ characterizing the `strength'
of the initial configuration and $\gamma$ is a `critical exponent',
whose value was found numerically to be $\gamma\approx 0.37$.
The critical value $p^*$ is distinguished by the fact that for
$p<p^*$ the space-time stays everywhere regular,
whereas for $p>p^*$ an apparent
horizon is formed signalling the formation of a black hole.
Subsequently the same type of critical behaviour was observed
in the collapse of an ideal liquid (`Radiation Fluid')
\cite{EvCol} and the collapse of axisymmetric gravitational wave packets
\cite{EvAb}. In both these cases the critical index $\gamma$
turned out to be close to the value
$\gamma\approx 0.37$ found for the scalar field collapse, suggesting
a kind of universality as known from critical behaviour in statistical
mechanics.

Obviously in an attempt to explain this behaviour a special
significance should be attributed to the critical solution
obtained for $p=p^*$. For the case of the `Radiation Fluid'
Evans and Coleman found that this solution is asymptotically
self-similar, i.e.\
invariant under a simultaneous rescaling of the radial
coordinate $r$ and the time $t$, whereas the critical solution
in the scalar field collapse exhibits only a discrete self-similarity
\cite{Chop}.
Spherically symmetric continuously self-similar solutions are
comparatively easy to analyze,
because the field equations can be reduced to
a system of ordinary differential equations. No such simplification
is obtained for solutions exhibiting only a discrete
self-similarity. In fact, Evans and Coleman where able to determine
the asymptotic form of the
critical solution in the case of the `Radiation Fluid'
integrating the
self-similar equations with certain boundary conditions to be
discussed subsequently.
These authors also proposed to determine the critical
index $\gamma$ through a linear stability analysis of the critical
solution. This is precisely the method we shall employ in the following.
Since only in the case of the fluid model we are able to determine the
asymptotic form of the
critical solution, we shall restrict the discussion to this case.

The considered Eulerian fluid
is described by the energy momentum tensor
$T_{\mu\nu}=(p+\rho)U_\mu U_\nu-pg_{\mu\nu}$ with energy density
$\rho$, pressure $p$ and 4-velocity $U_\mu$, adopting
the simple equation of state $p=k\rho$.
The `Radiation Fluid' considered in \cite{EvCol} corresponds to
the special case $k=1/3$.
The field equations were already given in \cite{EvCol}, however
we shall rewrite them in a form suitable for the discussion of the
self-similar solutions.
The spherically symmetric line element is parametrized as
\begin{equation}
ds^2={r^2\over A^2}\bigl({dt^2\over B^2t^2}-{dr^2\over r^2}\bigr)
-r^2d\Omega^2\;
\label{metric}
\end{equation}
and the 4-velocity is written as $U_{\hat t}=\sqrt{1-v^2}$,
$U_{\hat r}=v/\sqrt{1-v^2}$ in terms of the radial velocity $v$.
Instead of $\rho$ we use the dimensionless combination
$\tilde\rho\equiv\sqrt{4\pi}r\rho$.
In terms of the logarihmic variables
$\tau=\ln(-t)$ and $\sigma=\ln r-\ln(-t)$
the field equations become the autonomous sytem of PDE's
(with a prime denoting $d/d\sigma$ and a dot $d/d\tau$)
\begin{eqnarray}
&&\hspace{-1cm}
A'/A={1\over2}A^{-2}\bigl(1-A^2-2\tilde\rho{1+kv^2\over 1-v^2}\bigr)  \\
&&\hspace{-1cm}
B'/B=A^{-2}\bigl(2A^2-1+(1-k)\tilde\rho\bigr)        \\
&&\hspace{-1cm}
A'/A=(1+k){\tilde\rho v\over A^2B(1-v^2)}+\dot A/A   \\
&&\hspace{-1cm}
{1\over 1+k}(B+v)\tilde\rho'/\tilde\rho
+(1+Bv)v'/(1-v^2)=                       \nonumber     \\
&&(B+v)A'/A+vB'/B-{1+3k\over1+k}v+       \nonumber     \\
&&B\bigl({1\over1+k}\dot{\tilde\rho}/\tilde\rho+
v\dot v/(1-v^2)-\dot A/A\bigr)             \\
&&\hspace{-1cm}
{1\over 1+k}(k+(1+k)Bv+v^2)\tilde\rho'/\tilde\rho
+(B+2v+Bv^2)v'/(1-v^2)=                  \nonumber       \\
&&(1+2Bv+v^2)A'/A+(1+v^2)B'/B-{1-k+(1+3k)v^2\over1+k} \nonumber \\
&&+B\bigl(v\dot{\tilde\rho}/\tilde\rho
  +(1+v^2)\dot v/(1-v^2)-2v\dot A/A\bigr)
\label{field}
\end{eqnarray}
The equations for
self-similar solutions are obtained by dropping the terms with
$\tau$-derivatives. In this case the first and third equation
can be combined to yield the algebraic constraint
\begin{equation}
A^2=1-{2\tilde\rho\bigl(1+(1+k)v/B+kv^2\bigr)\over1-v^2}\;.
\label{constr}
\end{equation}
The general structure of the set of self-similar solutions has been
discussed in some detail by Bogoyavlenskii \cite{Bogo} and Ori and
Piran\cite{Ori}.
Solutions with a regular origin tend to the fixed point
$v=\tilde\rho=B=0$, $A=1$ for $\sigma\to -\infty$
and constitute a one-parameter family parametrized
by the central value $\rho_0$ of the density $\rho$.
The generic solution with a such a regular origin
turns out to develop a singularity (shock front) at the `sonic point'
for some finite value of $\sigma$.
Only for a discrete set of $\rho_0$-values
the formation of the shock front is avoided
and the solution stays regular for all values of $\sigma$.
Numerical analysis shows that these regular solutions can be labelled by
the number of zeros of the radial velocity field $v$. The solution
described by Evans and Coleman is the one with one zero of
$v$. Varying $k>0$ one finds that this type of solution exists only
up to a certain maximal value $k_{\rm max}\approx 0.888$, where
the singular sonic point changes its character.
As discussed in \cite{Bogo,Ori} there are three different cases
for the nature of this singular point.
It turns out that two of its eigenvalues degenerate for
$k=k_{\rm max}$ and the singular point changes from
an attractive node to an attractive focus.

One may ask why it is this regular self-similar solution that is of
relevance for the actual collapse. The reason may be that
the latter seems to proceed without the formation of a shock-front.
It is an interesting question, if
this behaviour is a consequence of the chosen class of initial
data or a more general feature due to the chosen equation of state.

The characteristic feature of solutions with initial data close
to those of the critical solution is, that they first approach the
latter, but eventually run away from it. In order to describe this
run-away we may employ a linear stability analysis as proposed
by Evans and Coleman\cite{EvCol}. In the linear approximation we expect
an unstable mode $\varphi$
of the critical solution with a time-dependence
$\varphi\sim \bar\varphi(\sigma,p)e^{-\omega\tau}$ with some
characteristic
frequency $\omega>0$. The requirement of regularity of $\varphi$
at the origin and the sonic point
acts as a boundary condition leading to a discrete frequency spectrum.
Since $\varphi$ vanishes for $p=p^*$, we may assume
$\bar\varphi(\sigma,p)=c(\sigma)(p-p^*)$.
Of particular relevance is the metric function $A$, which vanishes
at the position of an apparent horizon.
For $p$ close to $p^*$ it takes the form
\begin{equation}
A(\sigma,\tau,p)=A_{\rm ss}(\sigma)-
          c(\sigma)(p-p^*)e^{-\omega\tau}\;,
\label{pert}
\end{equation}
where $A_{\rm ss}$ denotes the self-similar back-ground.
Although the back-ground solution turns out to have
no apparent horizon (i.e. $A_{\rm ss}(\sigma)>0$),
the exponentially growing perturbation
will eventually lead to a zero of $A$ (assuming $p>p^*$ and $c>0$).
Here we have tacitly taken for granted
that the linear approximation is
still valid for perturbations of the same order as the background.
Let us hence assume $A(\sigma_{\rm h},\tau_{\rm h},p)=0$ for some values
$\tau_{\rm h}$,
$\sigma_{\rm h}$ determined by $A_{\rm ss}$ and $c$.
Using $r_{\rm h}=e^{\sigma_{\rm h}+\tau_{\rm h}}$ we obtain
\begin{equation}
A_{\rm ss}(\sigma_{\rm h})=c(\sigma_{\rm h})(p-p^*)
      {e^{\omega\sigma_{\rm h}}\over r_{\rm h}^\omega}\;
\label{app}
\end{equation}
leading to the desired relation
\begin{equation}
M_{\rm BH}=r_{\rm h}\sim (p-p^*)^{1\over\omega}\;.
\label{mass}
\end{equation}
For the critical index we read off the
expression $\gamma=1/\omega$.

\begin{table} [t]
\begin{center}
\begin{tabular}{|l|l||l|l|}  \hline
$k$        &$\gamma$&$k$& $\gamma$                 \\ \hline
0.01&0.1143      &  0.45&0.4400 \\
0.05&0.1478      &  0.50&0.4774 \\
0.10&0.1875      &  0.55&0.5159 \\
0.15&0.2251      &  0.60&0.5556 \\
0.20&0.2614      &  0.65&0.5967 \\
0.25&0.2970      &  0.70&0.6392 \\
0.30&0.3322      &  0.75&0.6834 \\
1/3 &0.3558      &  0.80&0.7294 \\
0.35&0.3676      &  0.85&0.7775 \\
0.40&0.4035      &  0.888&0.8157 \\
\hline
\end{tabular}
\end{center}
\caption[]{\label{tabgamma}
   Critical indices $\gamma$ for various values of $k$.}
\end{table}
Tab.~\ref{tabgamma} gives the numerically determined values of $\gamma$
for various values of $k$.
The value $\gamma=0.3558$ for $k=1/3$ compares very well with
value $\gamma\approx 0.36$ given by Evans and Coleman \cite{EvCol}.
The value $k=0.88$ is the maximal one for which a regular
self-similar solution could be found.
The limit $k\to 0$ is singular, since the sonic point merges
with the origin. Nevertheless $\gamma$ seems to approach a
non-vanishing limit.
As can be seen the critical index depends rather strongly on the value
of $k$, hence it is not universal.
It would be clearly desirable to check the predictions made for
$\gamma$ for values of $k$ different from $1/3$ by actual collapse
calculations.

{\bf Acknowledgement}\noindent
I am indebted to P.~Breitenlohner, B.~Br\"ugmann, F.~M\"uller-Hoissen
and A.~Rendall for stimulating discussions.

{\it Note added}: While this letter was in preparation
a paper \cite{Koike}
by T.~Koike, T.~Hara and S.~Adachi on the same subject
appeared.
Using a renormalization group type argument
the authors `explain' the approach to the critical solution
and final run-away of almost critical solutions. They obtain the
same expression for the critical exponent, which they determined
numerically for the special case $k=1/3$ of the `Radiation Fluid'.


\newpage
\begin{thebibliography}{9}
\bibitem{Chop}
M.W. Choptuik,
 {\it Phys. Rev. Lett.} {\bf 70} (1993) 9.

\bibitem{EvCol}
C.R. Evans and J.S. Coleman,
 {\it Phys. Rev. Lett.} {\bf 72} (1994) 1782.

\bibitem{Bogo}
O.I.Bogoyavlenskii, {\it Sov. Phys. JETP} {\bf 46} (1977) 633.

\bibitem{EvAb}
A.M. Abrahams and C.R. Evans,
 {\it Phys. Rev. Lett.} {\bf 70} (1993) 2980.

\bibitem{Ori}
A.Ori and T.Piran,
{\it Phys. Rev.} {\bf D42} (1990) 1068.

\bibitem{Koike}
T.~Koike, T.~Hara and S.~Adachi,
{\it Critical behaviour in gravitational collapse of radiation fluid
- A renormalization group (linear perturbation) analysis -},\\
gr-qc/9503007, TIT/HEP-284/COSMO-53.
%
\end{thebibliography}
\end{document}